\documentclass[twocolumn,aps,pre,groupedaddress]{revtex4-1}

\usepackage{amsmath}
\usepackage{graphicx}
\usepackage{amsmath}
\usepackage{graphicx}
\usepackage{subfig}
\usepackage{epstopdf}
\usepackage{amssymb}

\begin{document}
\title{Convergent Cross-Mapping and Pairwise Asymmetric Inference}
\author{James M. McCracken}
\email{jmccrac1@masonlive.gmu.edu}
\affiliation{School of Physics, Astronomy, and Computational Sciences \\ George Mason University \\ 4400 University Drive MS 3F3, Fairfax,VA. 22030-4444}
\author{Robert S. Weigel}
\email{rweigel@gmu.edu}
\affiliation{School of Physics, Astronomy, and Computational Sciences \\ George Mason University \\ 4400 University Drive MS 3F3, Fairfax,VA. 22030-4444}
\date{\today}

\begin{abstract}
Convergent Cross-Mapping (CCM) is a technique for computing specific kinds of correlations between sets of times series.  It was introduced by Sugihara {\em et al.\ }\cite{Sugihara2012} and is reported to be ``a necessary condition for causation'' capable of distinguishing causality from standard correlation.  We show that the relationships between CCM correlations proposed in \cite{Sugihara2012} do not, in general, agree with intuitive concepts of ``driving'', and as such, should not be considered indicative of causality.  It is shown that CCM causality analysis implies causality is a function of system parameters for simple linear and nonlinear systems.  For example, in a RL circuit, both voltage and current can be identified as the driver depending on the frequency of the source voltage.  It is shown that CCM causality analysis can, however, be modified to identify asymmetric relationships between pairs of time series that are consistent with intuition for the considered example systems for which CCM causality analysis provided non-intuitive driver identifications.  This modification of the CCM causality analysis is introduced as ``pairwise asymmetric inference'' (PAI) and examples of its use are presented.  
\end{abstract}

\maketitle

\section{Introduction}
Modern time series analysis includes techniques meant to discern ``driving'' relationships between different data sets.  These techniques have found application in a wide range of fields including neuroscience (e.g., \cite{Kaminski2001}), economics (e.g., \cite{dufour1998,dufour2006}), and climatology (e.g., \cite{mosedale2006}).  General casual relationships in time series data are also being studied in an effort to understand causality itself (e.g., \cite{eichler2012}).  

To date, most techniques for ``causal inference'' in time series fall into two broad categories: those related to transfer entropy and those related to Granger causality.  Transfer entropy (introduced in \cite{Schreiber2000}) and Granger causality (introduced in \cite{granger1969}) are known to be equivalent under certain conditions \cite{Barnett2009}.  In this article, we investigate a casual inference technique, called Convergent Cross-Mapping (CCM), that was recently introduced by Sugihara {\em et al.\ }\cite{Sugihara2012}; \footnote{Currently, there are no published equivalence conditions for CCM to either transfer entropy or Granger causality.}.

CCM is described as a technique that can be used to identify a necessary condition for causality between time series and is intended to be useful in situations where Granger causality is known to be invalid (i.e., in dynamic systems that are ``nonseperable'' \cite{Sugihara2012}). Granger causality is not causality as it is typically understood in physics \cite{Granger1980,liu2012,Roberts1985}.  We show that a similar conclusion can be made regarding CCM causality. 

CCM has been used to draw conclusions regarding the sardine-anchovy-temperature problem \cite{Sugihara2012}; confirm predictions of climate effects on sardines \cite{Deyle2013}; compare the driving effects of precipitation, temperature, and solar radiation on the atmospheric CO$_2$ growth rate \cite{Wang2014}; study the driving relationship between pressure and displacement of abdominal parts in insects \cite{Bozorgmagham2013}; and to quantify cognitive control in developmental psychology \cite{Anastas2013}.  The wide range of applications already appearing for the relatively new CCM technique is a testament to the importance of time series causality studies.  This work presents examples in which CCM does not provide consistent qualification of an intuitive notion of causality.  (However, the domain of applicability of CCM remains an open question; i.e., the method may have worked as expected in the above-cited papers despite its apparent failure in the examples presented in this article.) 

We begin with a review of the work of Sugihara {\em et al.} \cite{Sugihara2012}, including an extended evaluation of the coupled logistic map example.  After showing examples where CCM analysis gives results that are inconsistent with intuitive notions of driving, we introduce ``pairwise asymmetric inference'' (PAI) and show that it can be used to identify asymmetric relationships that are consistent with intuitive notions of driving.

\section{Convergent Cross-Mapping}
CCM is closely related to simplex projection \cite{Sugihara1990,Sugihara1990a}, which predicts a point in the times series $X$ at a time $t+1$, labeled $X_{t+1}$, by using the points with the most similar histories to $X_t$.  Similarly, CCM uses points with the most similar histories to $X_t$ to estimate $Y_t$.  The CCM correlation is the squared Pearson correlation coefficient \footnote{This definition differs slightly from the definition in \cite{Sugihara2012}, which uses the un-squared Pearson’s correlation coefficient.  We use the square of this value to avoid dealing with negative correlation values.  This subtle change in the definition does not affect the conclusions drawn in \cite{Sugihara2012}, as can be seen in our reproduction of key plots from that work: Figure \ref{fig:BGridPlot} and Figure \ref{fig:Sug3CDredo}.} between the original time series $Y$ and an estimate of $Y$ made using its convergent cross-mapping with $X$, which is labeled as $Y|\tilde{\mathbf{X}}$:
$$
C_{YX} = \left[\rho(Y,Y|\tilde{\mathbf{X}})\right]^2\;\;.
$$
Any pair of times series, $X$ and $Y$, will have two CCM correlations, $C_{YX}$ and $C_{XY}$, which are compared to determine the CCM causality.  For example, Sugihara {\em et al.\ }\cite{Sugihara2012} define a difference of CCM correlations
\begin{equation}
\label{eqn:delta}
\Delta = C_{YX} - C_{XY}
\end{equation}
and use the sign of $\Delta$ to determine the CCM causality between $X$ and $Y$.

If $X$ can be estimated using $Y$ better than $Y$ can be estimated using $X$ (e.g., if $\Delta < 0$), then $X$ is said to ``CCM cause'' $Y$.

\subsection{CCM Algorithm}
\label{sec:appA}
The CCM algorithm may be written in terms of five steps:
\begin{enumerate}
\item 
Create the {\em shadow manifold} for $X$, called $\tilde{\mathbf{X}}$;
\item Find the nearest neighbors to a point in the shadow manifold at time $t$, which is labeled $\tilde{\mathbf{X}}_t$;
\item Create weights using the nearest neighbors;
\item Estimate $Y$ using the weights; (this estimate is called $Y|\tilde{\mathbf{X}}$); and
\item Compute the correlation between $Y$ and $Y|\tilde{\mathbf{X}}$. 
\end{enumerate}
The steps are described in more detail below.  

\subsubsection{Create Shadow Manifold $\tilde{\mathbf{X}}$}
\label{sec:shadow}
Given an embedding dimension $E$, the {\em shadow manifold} of $X$, called $\tilde{\mathbf{X}}$, is created by associating an $E$-dimensional vector (also called a {\em delay vector}) to each point $X_t$ in $X$, i.e., $\tilde{\mathbf{X}}_t=\left(X_t,X_{t-\tau},X_{t-2\tau},\ldots,X_{t-(E-1)\tau}\right)$.  The first such vector is created at $t=1+(E-1)\tau$ and the last is at $t=L$ where $L$ is the number of points in the time series (also called the {\em library length}).  

\subsubsection{Find Nearest Neighbors}
The minimum number of points required for a bounding simplex in an $E$-dimensional space is $E+1$ \cite{Sugihara1990,Sugihara1990a}.  Thus,  the set of $E+1$ nearest neighbors must be found for each shadow manifold $\tilde{\mathbf{X}}_t$.  For each $\tilde{\mathbf{X}}_t$, the nearest neighbor search results in a set of distances that are ordered by closeness $\{d_1,d_2,\ldots,d_{E+1}\}$ and an associated set of times $\{\hat{t}_1,\hat{t}_2,\ldots,\hat{t}_{E+1}\}$.  The distances from $\tilde{\mathbf{X}}_t$ are
$$
d_i = D\left(\tilde{\mathbf{X}}_t,\tilde{\mathbf{X}}_{\hat{t}_i}\right)\;\;,
$$
where $D(\mathbf{a},\mathbf{b})$ is the Euclidean distance between vectors $\mathbf{a}$ and $\mathbf{b}$.

\subsubsection{Create Weights}
Each of the $E+1$ nearest neighbors are be used to compute an associated weight.  The weights are defined as
$$
w_i = \frac{u_i}{N}\;\;,
$$
where
$u_i = e^{-d_i/d_1}$ and the normalization factor is $N = \sum_{j=1}^{E+1} u_j\;\;.$

\subsubsection{Find $Y|\tilde{\mathbf{X}}$}
A point $Y_t$ in $Y$ is estimated using the weights calculated above.  This estimate is
$$
Y_t|\tilde{\mathbf{X}} = \sum_{i=1}^{E+1} w_i Y_{\hat{t_i}}\;\;.
$$

\subsubsection{Compute the Correlation}
The CCM correlation is defined as 
$$
C_{YX} = \left[\rho\left(Y,Y|\tilde{\mathbf{X}}\right)\right]^2\;\;,
$$
where $\rho\left(A,B\right)$ is the standard Pearson's correlation coefficient between $A$ and $B$.  

The CCM algorithm depends on the embedding dimension $E$ and the lag time step $\tau$.  A dependence on $E$ and $\tau$ is a feature of most state space reconstruction (SSR) methods \cite{Hong2006,vlachos2009,Small2004}, so an $E$ and $\tau$ dependence is not unexpected.  Sugihara {\em et al.} mention that ``optimal embedding dimensions'' are found using univariate SSR \cite{Sugihara2012} (supplementary material), and other methods for determining $E$ and $\tau$ for SSR algorithms can be found in the literature (e.g., \cite{Hong2006,Small2004,Kennel1992}).

\subsection{CCM Example}
\label{sec:2Pop}
Consider the example system used by Sugihara {\em et al.\ }\cite{Sugihara2012}:
\begin{eqnarray}
\label{eqn:2pop}
X_t &=& X_{t-1}\left(r_x-r_x X_{t-1}-\beta_{xy} Y_{t-1}\right)\\
Y_t &=& Y_{t-1}\left(r_y-r_y Y_{t-1}-\beta_{yx} X_{t-1}\right)
\end{eqnarray}
where the parameters $r_x,r_y,\beta_{xy},\beta_{yx}\in\mathbb{R}\ge 0$.  This pair of equations is a specific form of the two-dimensional coupled logistic map system \cite{Lloyd1995}.

\begin{figure}[ht]
\begin{tabular}{l}
\includegraphics[scale=0.5]{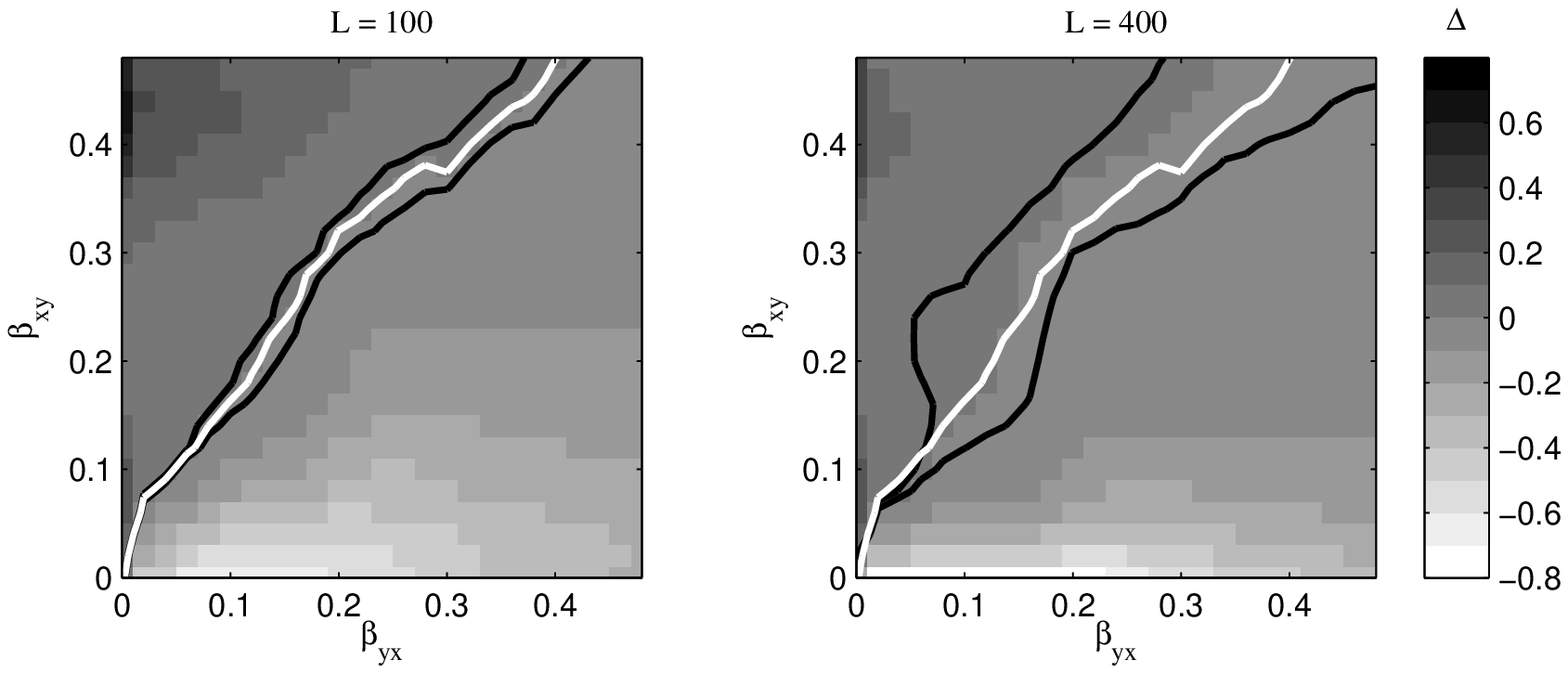}\\
\includegraphics[scale=0.5]{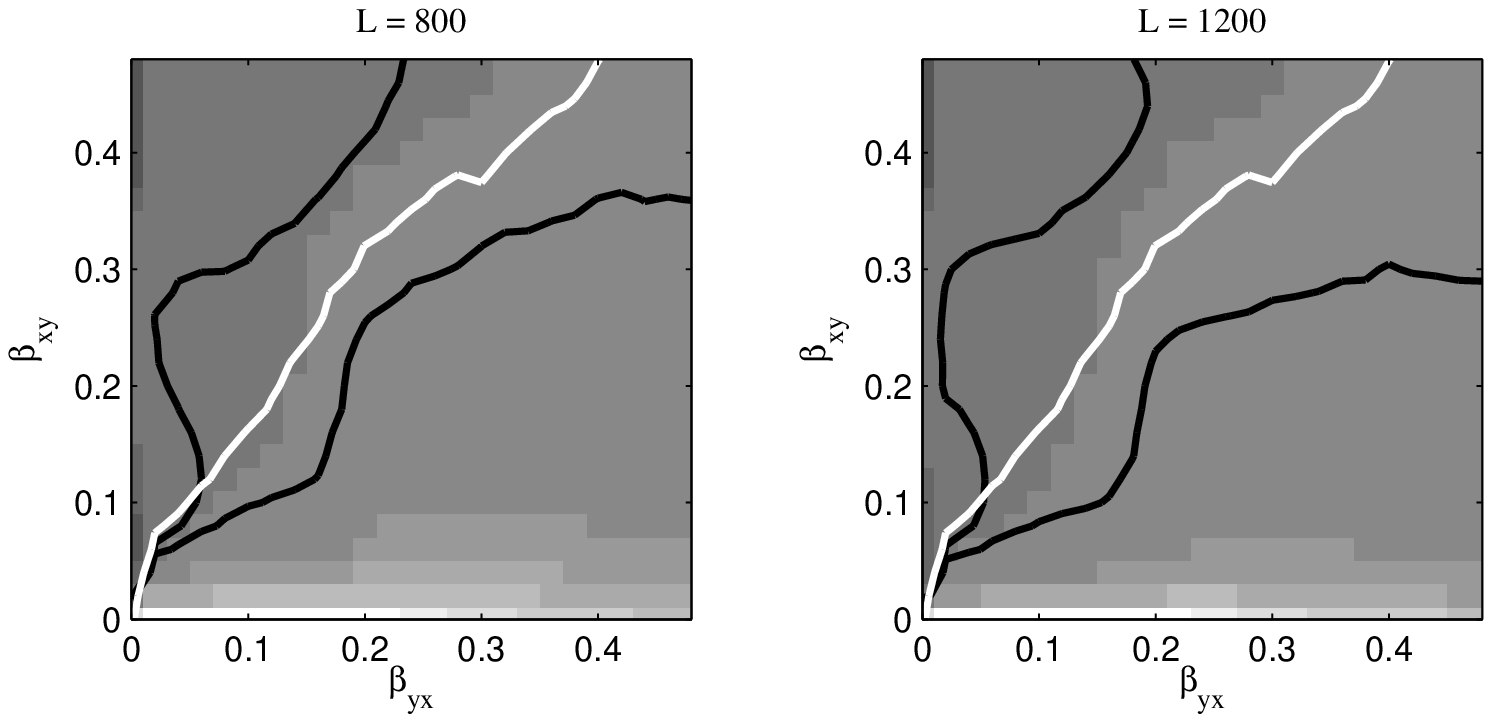}
\end{tabular}
\caption{The dependence of Eqn.\ \ref{eqn:delta} on $\beta_{xy}$ and $\beta_{yx}$.  The white line is a contour for $\Delta=0$; Black lines are contours for $\Delta=\pm 0.01$.}
\label{fig:BGridPlot}
\end{figure}

In this example, the CCM causality relationship between $X$ and $Y$ is determined using a sampling of both the initial conditions and the system parameters, calculating $\Delta$, and demonstrating the necessary convergence.  The term {\em convergence} is used here in the same sense as it was used in \cite{Sugihara2012}; i.e., ``convergence means that cross-mapped estimates improve in estimation skill with time-series length $L$'' \footnote{In terms of $\Delta$, convergence to a value $\kappa$ means $\lim_{L\rightarrow\infty}|\Delta-\kappa| = 0$.  As the ``estimation skill'' of the CCM algorithm increases, each of the CCM correlations, $C_{XY}$ and $C_{YX}$, converge to some fixed value.  Thus, $\Delta$ converges to some fixed value $\kappa$.  $\Delta$ depends on the library length $L$ because the CCM algorithm depends on $L$.  However, the complicated algorithmic dependence makes formally solving this limit difficult.  As such, convergence is determined from plots, following the method used in  \cite{Sugihara2012}.}.  The dynamic parameters $r_x$ and $r_y$ are sampled from normal distributions $\mathcal{N}\left(\mu_{rx},\sigma_{rx}\right)$ and $\mathcal{N}\left(\mu_{ry},\sigma_{ry}\right)$, respectively.  The initial conditions $X_0$ and $Y_0$ are also sampled from normal distributions, specifically $\mathcal{N}\left(\mu_{x0},\sigma_{x0}\right)$ and $\mathcal{N}\left(\mu_{y0},\sigma_{y0}\right)$.  The coupling parameters $\beta_{xy}$ and $\beta_{yx}$ are then varied over the interval $[10^{-6},1]$ in steps of 0.02 to produce the plots seen in Figure \ref{fig:BGridPlot}.

Sugihara {\em et al.\ }consider convergence to be critically important for determining CCM causality, and note that it is ``a key property that distinguishes causation from simple correlation'' \cite{Sugihara2012}.  Figure \ref{fig:BGridPlot} shows plots created with several different library lengths to illustrate the convergence of $\Delta$ for this example.  Typically, for convenience, the (approximately) converged CCM correlation values will be reported and proof of convergence will be implied, rather than shown.

The idea is that $\beta_{xy}>\beta_{yx}$ intuitively implies $Y$ ``drives'' $X$ more than $X$ ``drives'' $Y$.  Stated more formally, $\beta_{xy}>\beta_{yx}\Rightarrow\Delta>0$, which is reported as ``$Y$ CCM causes $X$''.  Likewise, $\beta_{xy}<\beta_{yx}$ implies $X$ CCM causes $Y$ and $\beta_{xy}=\beta_{yx}$ implies no CCM causality in the system.  It will be shown below that CCM causality is not necessarily related to causality as it is typically understood in physics.

\section{Simple Example Systems}
The usefulness of the CCM algorithm in identifying drivers among sets of time series can be explored by using simple example systems.  Each of the following examples intuitively supports the conclusion that $X$ drives $Y$, and CCM analysis (with $E=3$ and $\tau=1$) yields values of $\Delta$ that support conclusions that do not agree with intuition for all parameter choices.

\subsection{Linear Example}
\begin{figure}[ht]
\begin{tabular}{l}
\includegraphics[scale=0.8]{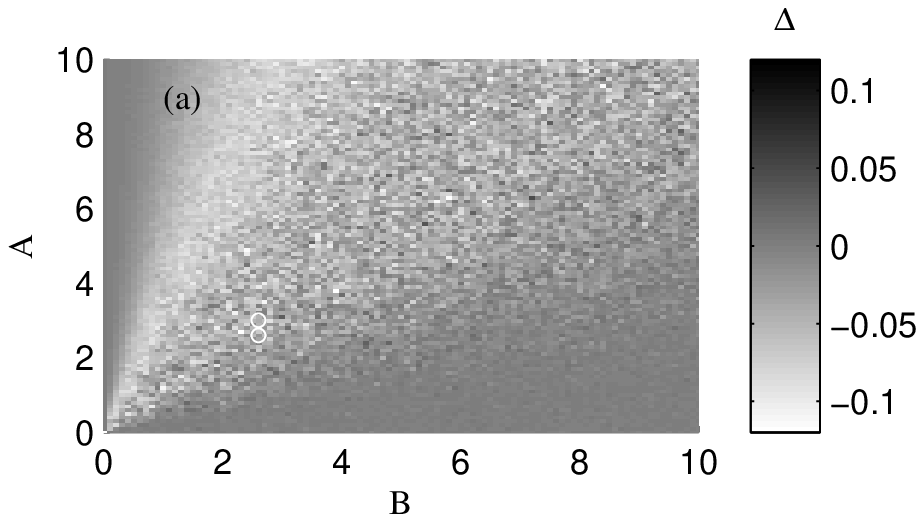} \\
\includegraphics[scale=0.8]{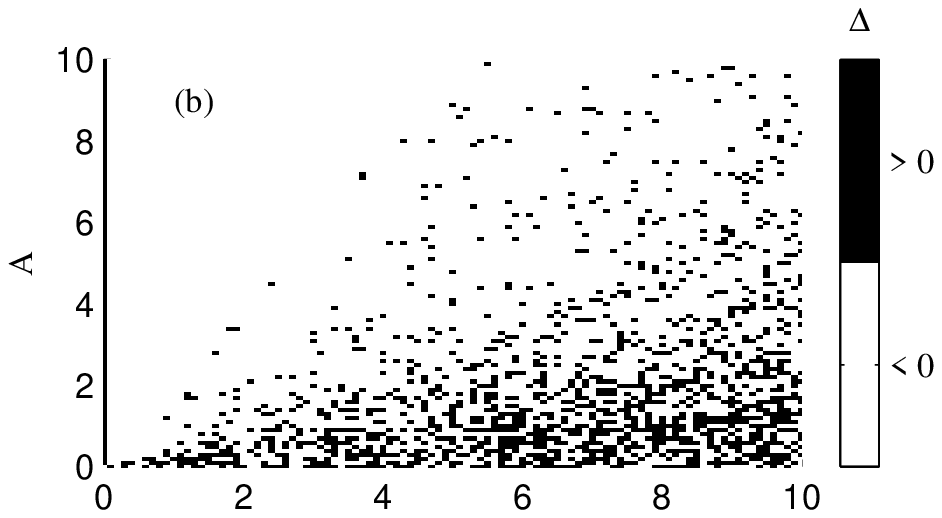} \\
\includegraphics[scale=0.8]{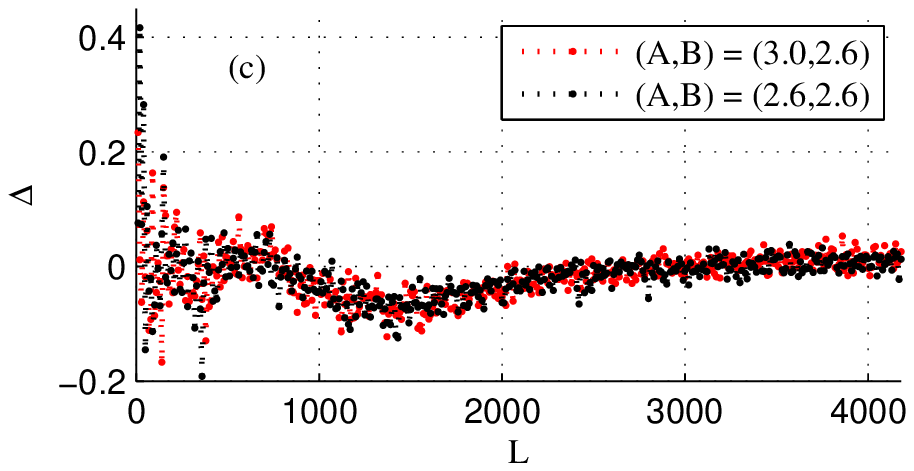} \\
\end{tabular}
\caption{Dependence of the sign of $\Delta$ from Eqn.\ \ref{eqn:delta} on $A$ and $B$. (a) $\Delta$ is calculated as described in the text; (b) Two-color version of (a) where black indicates $\Delta>0$, i.e.,\ $Y$ CCM causes $X$, and white indicates $\Delta<0$, i.e.,\ $X$ CCM causes $Y$; (c) The dependence of $\Delta$ on library length $L$ is shown for points $(A,B) = (2.6,2.6)$ and $(A,B)=(3.0,2.6)$ marked in (a) with white circles.}
\label{fig:linearex1}
\end{figure}

Consider the linear example dynamical system of
\begin{eqnarray}
\label{eq:linearex}
X_t &=& \sin(t)\\
Y_t &=& AX_{t-1}+B\eta_t,
\end{eqnarray}
with $A,B\in\mathbb{R}\ge 0$ and $\eta_t\sim\mathcal{N}\left(0,1\right)$.  Specifically, consider $A,B\in[0,10]$ in increments of 0.1.  Figure \ref{fig:linearex1} shows $\Delta$ for this example given a library length of $L=2000$.

Figure \ref{fig:linearex1} (c) shows (for $(A,B) = (2.6,2.6)$ and $(A,B)=(3.0,2.6)$) that $\Delta$ is more negative at shorter library lengths but converges to a point near zero as the library length is increased.  The convergence of CCM correlations is emphasized \cite{Sugihara2012}, so the seemingly counter-intuitive behavior of $\Delta$ (and $C_{XY}$ and $C_{YX}$) in Figure \ref{fig:linearex1} implies that the CCM correlations may not be a reliable measure of ``driving'' (at least not the intuitive definition) for this simple linear example system.

The expected conclusion that $X$ drives $Y$, corresponding to $X$ CCM causes $Y$ requires $\Delta<0$.  But, it can be seen from Figure \ref{fig:linearex1} (b), the sign of $\Delta$ depends on $A$ and $B$.  Given that the intuitive conclusion of $X$ drives $Y$ in Eqn.\ \ref{eq:linearex} does not depend on $A$ and $B$, it would seem that $\Delta$ does not reliably reflect the intuitive conclusion in this linear example system.  

\subsection{Non-Linear Example}
\begin{figure}[ht]
\includegraphics[scale=0.5]{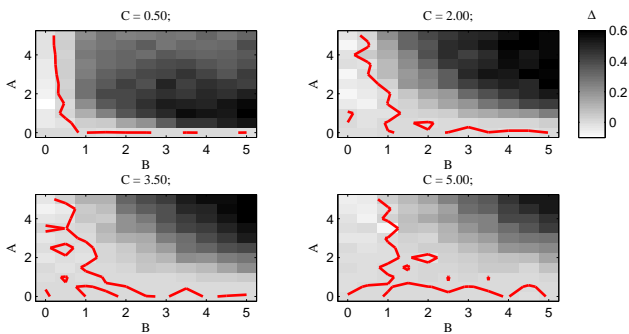} \\
\caption{The sign of $\Delta$, and thus the CCM causality, depends on $A$, $B$, and $C$. Contour lines indicate where $\Delta=0$.}
\label{fig:nonlinearex}
\end{figure}

Consider the non-linear dynamical system of
\begin{eqnarray}
\label{eqn:nonlinearEX}
X_t &=& \sin(t)\\
Y_t &=& AX_{t-1}\left(1-BX_{t-1}\right)+C\eta_t,
\end{eqnarray}
with $A,B,C\in\mathbb{R}\ge 0$ and $\eta_t\sim\mathcal{N}\left(0,1\right)$.  Specifically, consider $A,B,C\in[0,5]$ in increments of 0.5.  Figure \ref{fig:nonlinearex} shows $\Delta$ for specific values of $C$ for a library length of $L=2000$.
As in the previous example, the expectation for this system is that $\Delta$ should be negative, independent of the parameters $A$, $B$, and $C$.  However, it can be seen from the plots that the sign of $\Delta$ can depend on all three parameters.  Thus, this simple non-linear example leads to a conclusion similar to the previous linear example; i.e., $\Delta$ does not reliably reflect intuitive notions of driving.

\subsection{RL Circuit Example}
\label{sec:rlcirc}
Both of the previous examples included a noise term, $\eta_t$.  The failure of CCM analysis to give expected conclusions about the drivers in the previous examples may be due to a limitation of the algorithm with respect to noise.  This can be investigated by considering a system without noise.  Consider a series circuit containing a resistor, inductor, and time varying voltage source related by
\begin{equation}
\label{eqn:it}
\frac{dI}{dt} = \frac{V(t)}{L} - \frac{R}{L} I,
\end{equation}
where $I$ is the current at time $t$, $V(t)= \sin\left(\Omega t\right)$ is the voltage at time $t$, $R$ is the resistance, and $L$ is the inductance.  Eqn. \ref{eqn:it} was solved using the {\em ode45} integration function in MATLAB.  The time series $V(t)$ is created by defining values at fixed points and using linear interpolation to find the time steps required by the ODE solver.  

Consider the situation where $L=10$ Henries and $R=5$ Ohms are constant.  Physical intuition is that $V$ drives $I$, and so we expect to find that $V$ CCM causes $I$ (i.e., $C_{VI}>C_{IV}$ or $\Delta = C_{VI}-C_{IV} > 0$). 

\begin{figure}[ht]
\includegraphics[scale=0.9]{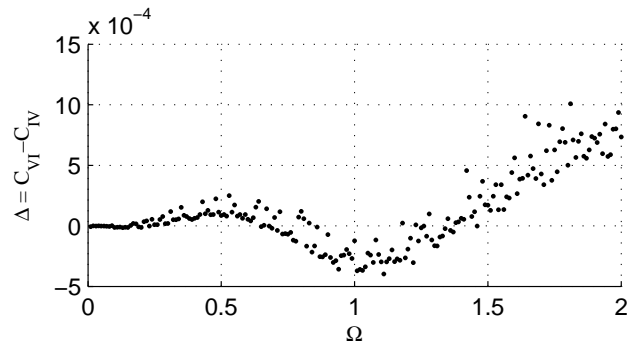} \\
\caption{$\Delta$ dependence on $\Omega$ for a series RL circuit with a sinusoidal voltage source.  The sign of $\Delta$ implies $V$ CCM causes $I$ and $I$ CCM causes $V$ depending on the voltage frequency.}
\label{fig:Av}
\end{figure}

Consider evaluating the CCM correlations $C_{VI}$ and $C_{IV}$ for each $\Omega\in[0.01,2.0]$ in steps of $0.01$.  The CCM correlations are found using $E=2$ and $\tau=1$ and are used to calculate $\Delta = C_{VI}-C_{IV}$, which is plotted in Figure \ref{fig:Av}.
$\Delta$ does not consistently agree with intuition in this example either.  Changing the embedding dimension, $E$, used to calculate $\Delta$ leads to plots that are different than Figure \ref{fig:Av}, but in all of the cases tested (i.e., $E=2,3,4$), the sign of $\Delta$ changes over the domain $\Omega\in[0.1,1.5]$. 

The resistance and inductance of the circuit are fixed and the voltage is varied from $1\times 10^{-2}$ to $2.0$ volts in discrete steps of $0.01$ volts as described by Eqn.\ \ref{eqn:it}.  Physically changing the voltage and witnessing a resulting change in the current is enough to convince most people that the voltage ``drives'' the current.  Rigorous statistical hypothesis testing can be performed to strengthen the confidence in such a conclusion.  Yet, from Figure \ref{fig:Av}, the voltage does not consistently ``CCM cause'' the current as $\Omega$ is changed. 

It may be argued that the relatively small values (as compared to the previous examples) of $\Delta$ plotted in Figure \ref{fig:Av} indicate that the correct conclusion should be either (1) there is no CCM causality in the system or (2) CCM causality is not applicable to this system.  However, conclusion (1) conflicts with the intuitive notion of an RL circuit as a strongly driven system and conclusion (2) conflicts with identifying CCM causality as a general qualifier of ``driving'' in dynamical systems.

\section{Pairwise Asymmetric Inference}
\label{sec:PAI}

\begin{figure}[ht]
\includegraphics[scale=0.5]{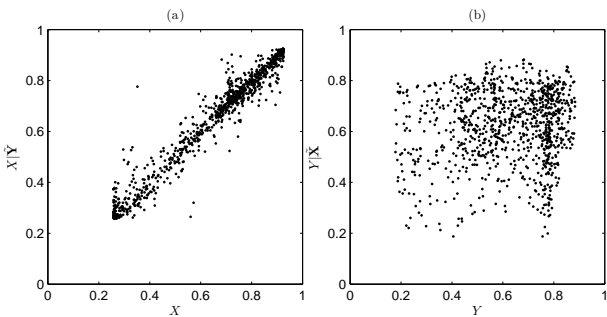} 
\caption{Correlation plots between a given time series and its convergent cross-mapped estimate.  (a) Reproduction of Figures 3C from \cite{Sugihara2012}. (b) Reproduction of Figures 3D from \cite{Sugihara2012}.}
\label{fig:Sug3CDredo}
\end{figure}

Consider the example system of Eqn.\ \ref{eqn:2pop} with $r_y=r_y=3.7$, $X_0 = 0.2$, $Y_0=0.4$, $\beta_{xy}=0$, and $\beta_{yx}=0.32$.  These parameters correspond to Figure 3C and D of \cite{Sugihara2012} (with $E=2$, $\tau=1$, and $L=1000$).  Plots of the correlation between $X$ and $X|\tilde{\mathbf{Y}}$ (i.e., $X$ estimated using the weights found from the shadow manifold of $Y$), as well as, $Y$ and $Y|\tilde{\mathbf{X}}$ are shown in Figure \ref{fig:Sug3CDredo}.
This leads to $\Delta=C_{YX}-C_{XY}\approx 0.11 - 0.97 = -0.86 <0$, which implies $X$ CCM causes $Y$.  This result agrees with intuition because $\beta_{xy}=0 < \beta_{yx} = 0.32$.

\begin{figure}[ht]
\begin{tabular}{c}
\includegraphics[scale=0.5]{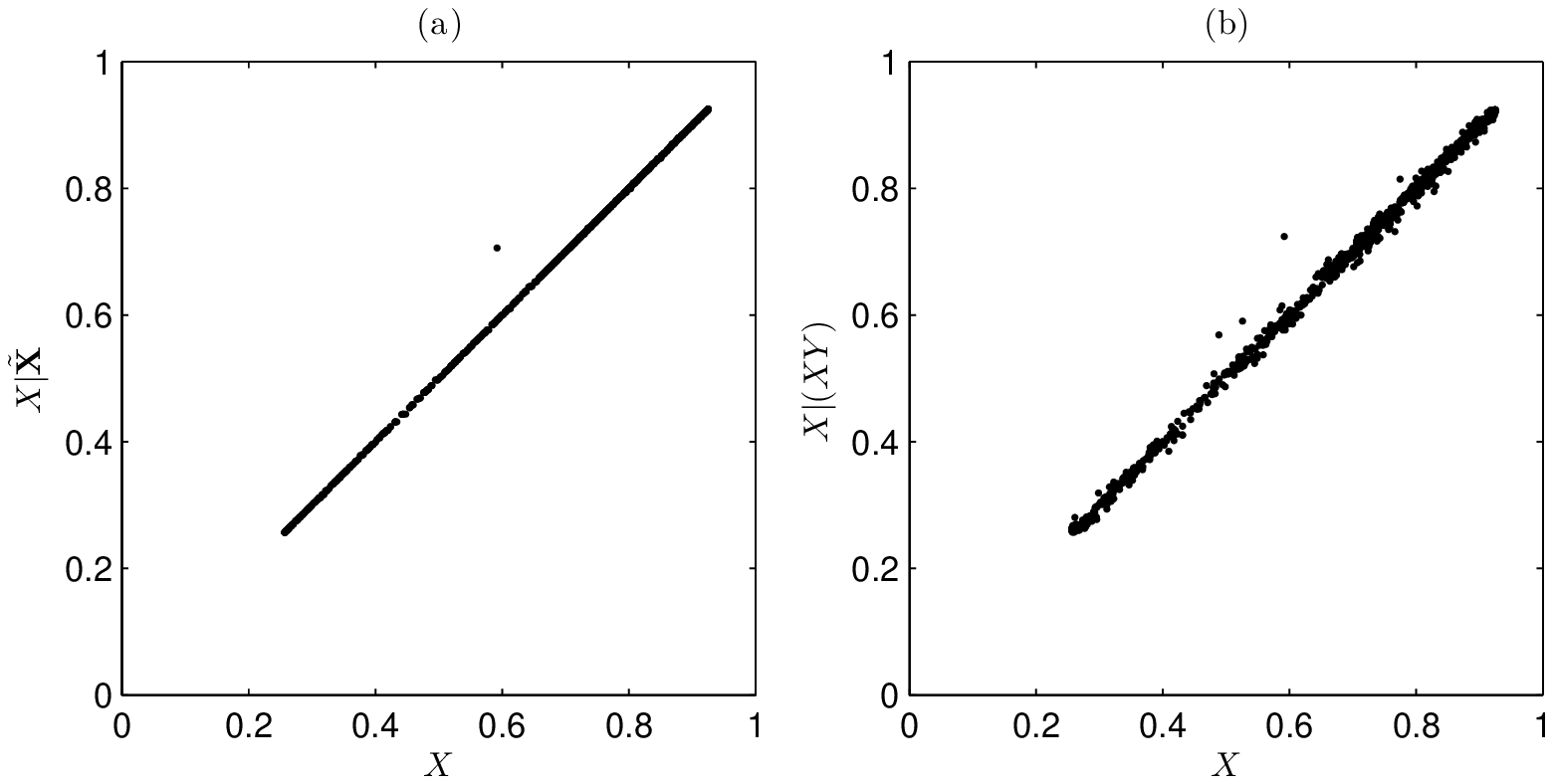} \\
\includegraphics[scale=0.5]{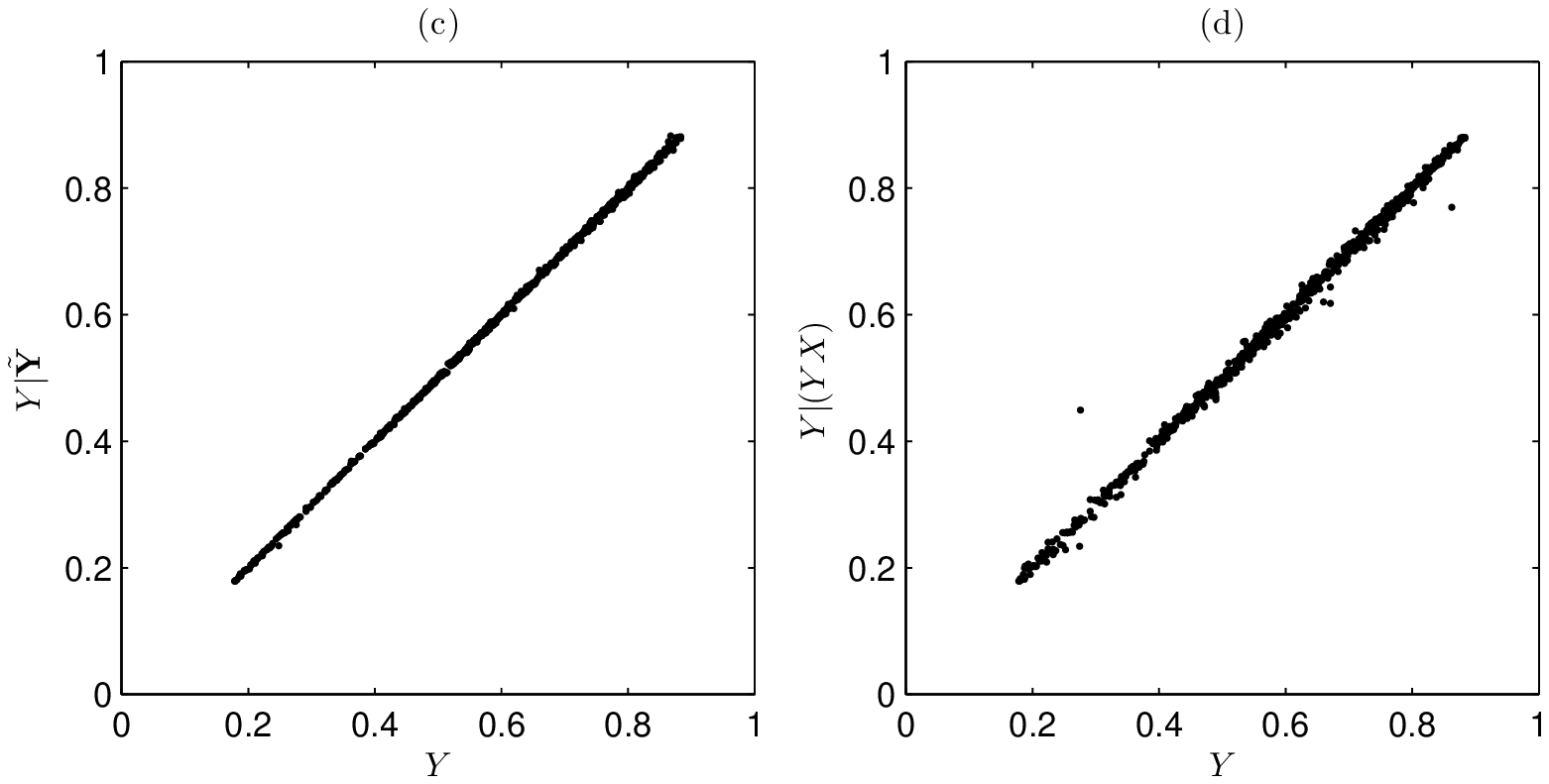}
\end{tabular}
\caption{Stronger correlations, as compared to Figure \ref{fig:Sug3CDredo}, can be seen between a time series and its estimate when the shadow manifold includes points from the time series it is estimating.}
\label{fig:PAIintro}
\end{figure}

The correlations shown in Figure \ref{fig:Sug3CDredo} are not the only correlations that can be tested.  Consider, for example, the correlation between $X$ and the corresponding $X|\mathbf{X}$, which is estimated using weights found from the shadow manifold of $X$ itself.  The time series $X$ may also be estimated using a multivariate shadow manifold consisting of points from both $X$ and $Y$ \cite{Deyle2013}.  For example, an $E+1$ dimensional point in the a multivariate shadow manifold constructed using both $X$ and $Y$ may be defined as $\tilde{\mathbf{X}}_t=(X_t,X_{t-\tau},X_{t-2\tau},\ldots,X_{t-(E-1)\tau},Y_t)$.  An estimate of $X$ using weights from a shadow manifold using this specific construction will be referred to as $X|(XY)$ and the correlation between this estimate and the original time series will be labeled $C_{X(XY)}$.  See Figure \ref{fig:PAIintro}.

A difference in CCM correlations similar to $\Delta$ can be defined using the multivariate embedding.  Consider $\Delta^\prime = C_{Y(YX)} - C_{X(XY)}$.  It might be argued, in close parallel to the arguments given in \cite{Sugihara2012} for $\Delta$, that an intuitive definition of ``driving'' might be captured by the sign of $\Delta^\prime$.  For example, if $\Delta^\prime<0$, then the single time step of $Y$ added to the delay vectors constructed from $X$ create stronger estimators of $X$ than the single time step of $X$ added to the delay vectors constructed from $Y$ do for $Y$.  Thus, it might be argued, that $Y$ contains more ``information'' about $X$, which leads to the conclusion $X$ drives $Y$.  The example system and parameters (including $E$, $\tau$, and $L$) described at the beginning of this section leads to $\Delta^\prime \approx -3\times 10^{-4}$ which agrees with the previously discussed conclusions of ``$X$ CCM causes $Y$'' and ``$X$ drives $Y$''.  Using the multivariate embedding described above to explore ``driving'' relationships between pairs of time series will be referred to as {\em pairwise asymmetric inference} or {\em PAI}.  

\begin{figure}[ht]
\begin{tabular}{c}
\includegraphics[scale=0.8]{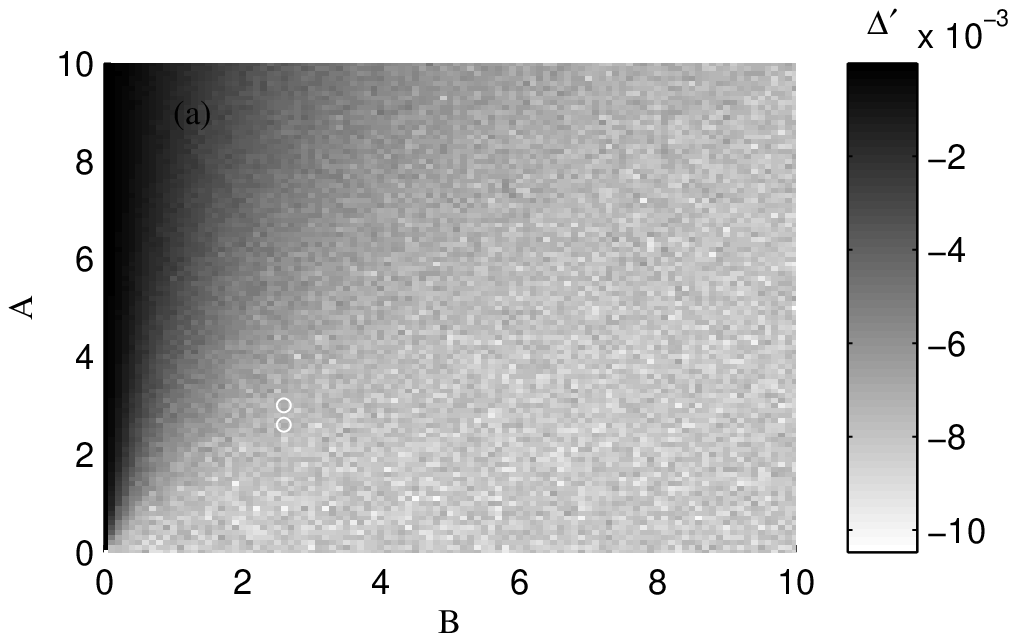} \\
\includegraphics[scale=0.8]{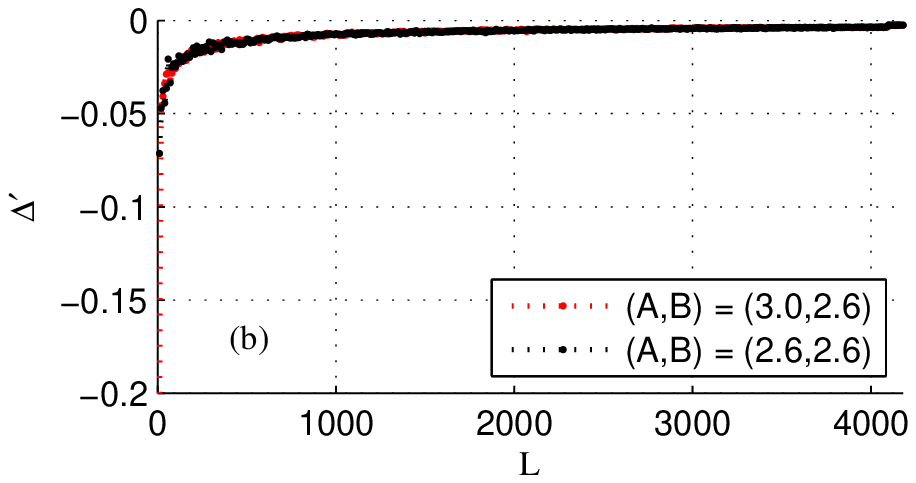} \\
\end{tabular}
\caption{(a) Reproducing Figure \ref{fig:linearex1} (a) using PAI rather than CCM.  $\Delta^\prime<0\;\forall A,B$ implying $X$ ``PAI drives'' $Y$. (b) Reproducing Figure \ref{fig:linearex1} (c) using PAI rather than CCM.  $\Delta^\prime$ does not display the apparent erratic behavior seen in $\Delta$ in Figure \ref{fig:linearex1}.}
\label{fig:linearExPAI}
\end{figure}

Consider a comparison of PAI and CCM given the linear example system from above, i.e., Eqn.\ \ref{eq:linearex}.  Figure \ref{fig:linearExPAI} shows $\Delta^\prime$ as a function of $A$ and $B$ using of the same $E$, $\tau$, $L$, and step sizes as was used to produce Figure \ref{fig:linearex1}.
$\Delta^\prime<0\;\forall A,B$ in the domains shown in the figure.  Thus, the sign of $\Delta^\prime$ is in agreement with an intuitive notion of driving more consistently than $\Delta$.  $\Delta^\prime$ is significantly smaller than $\Delta$, which is expected since the correlation of $X$ and $Y$ with their ``self estimation'' counterparts of $X|\tilde{\mathbf{X}}$ and $Y|\tilde{\mathbf{Y}}$ are initially very high, even without the multivariate additions.  But, if the concept of driving is determined solely on the sign of $\Delta^\prime$, then, at least for the simple linear example presented here, PAI is a consistent qualifier of ``driving''.

Reproducing Figure \ref{fig:linearex1} (c) using PAI shows an apparent reduction in some of the erratic behavior seen in CCM.  See Figure \ref{fig:linearExPAI}.

\begin{figure}[ht]
\includegraphics[scale=0.5]{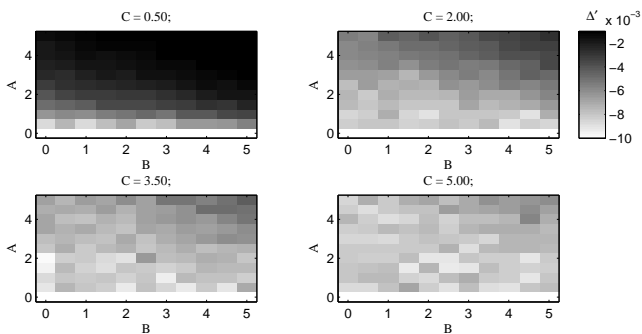} \\
\caption{Reproducing Figure \ref{fig:nonlinearex} using PAI rather than CCM.  $\Delta^\prime<0\;\forall A,B,C$ implying $X$ ``PAI drives'' $Y$ consistently in the plotted parameter domains.}
\label{fig:nonlinearEXPAI}
\end{figure}

The conclusions that PAI agrees with intuition more consistently than CCM is also supported by the non-linear example system, Eqn.\ \ref{eqn:nonlinearEX}.  Figure \ref{fig:nonlinearEXPAI} plots $\Delta^\prime$ as a function of $A$, $B$ and $C$ using of the same $E$, $\tau$, $L$, and step sizes that were used to produce Figure \ref{fig:nonlinearEXPAI}.
Again in contrast to the CCM figure, PAI agrees with intuition for all the plotted values of $A$, $B$, and $C$ (i.e.\ $\Delta^\prime<0\;\forall A,B,C$ in the domains shown).

\begin{figure}[ht]
\includegraphics[scale=0.8]{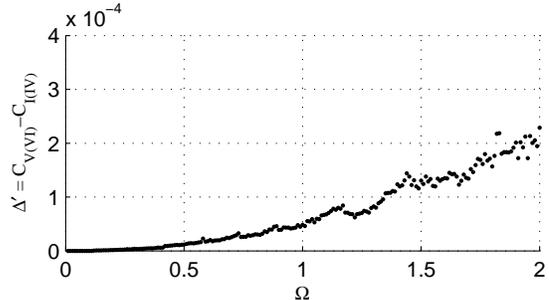} \\
\caption{Reproducing Figure \ref{fig:Av} using PAI rather than CCM.  $\Delta^\prime<0\;\forall \Omega$ implying $V$ ``PAI drives'' $I$ consistently across the plotted domain for $\Omega$.}
\label{fig:AvPAI}
\end{figure}

Finally, a comparison of PAI and CCM for the RL circuit example leads to similar conclusions.  The expectation is the $V$ drives $I$; thus, it is expected that $V$ PAI drives $I$ which implies $\Delta^\prime = C_{V(VI)} - C_{I(IV)} > 0$ (which is what is observed).  See Figure \ref{fig:AvPAI}.

\section{Conclusion}

In this work we have shown that the recently introduced and frequently used Convergent Cross Mapping (CCM) method can lead to conclusions about a driver in a system that does not agree with intuition and the identified driver can depend on system parameters.  For the examples presented in this article, PAI better indicates ``driving'' relationships that both agree with intuition and are consistent in the sense that the driver identified does not depend on system parameters.  

The introduced Pairwise Asymmetric Inference method (PAI) attempts to keep the model-independent benefits of CCM while making it more robust. (SSR methods such as CCM and PAI are model-independent, which may be seen as a benefit over Granger causality methods.) PAI may be useful exploratory data analysis.  For example, PAI may help guide the development of physical causality models (e.g., by suggesting future experiments) in scenarios involving a large collection of simultaneous time series measurements of different variables in a system for which no {\em a priori} notions of causality in the system exist. 

The given definition of $\Delta^\prime$ in PAI, the sign of which is used to identify a dominant driver, is not without its own difficulties, however.  For example, $\Delta^\prime$ does not account for the differences between correlations between $X$ and $X|\tilde{\mathbf{X}}$ and $Y$ and $Y|\tilde{\mathbf{Y}}$.  Such differences may bias conclusions drawn from using $\Delta^\prime$ without proper care.  As a concrete example, consider the example system and parameters (including $E$, $\tau$, and $L$) described at the beginning of Section \ref{sec:PAI}.  The value $\Delta^\prime \approx -3\times 10^{-4}$ was already discussed, but notice $C_{YY}-C_{XX} \approx 1.5\times10^{-3}$, indicating that $Y$ is a better ``self estimator'' than $X$ (though both $C_{YY},C_{XX}>0.99$).  How does this fact affect interpretations of the $\Delta^\prime<0$ result, which was that $X$ PAI drives $Y$?  Such questions are still open.  It may be argued that a different measure may be more suitable, such as $\Delta^{\prime\prime} = |C_{Y(YX)}-C_{YY}|-|C_{X(XY)}-C_{XX}|$.  For this example, $\Delta^{\prime\prime} \approx 3.9\times 10^{-4}$, which does not agree with intuition, despite the agreement of both $\Delta$ and $\Delta^\prime$.  There are still many open questions in the study of driving relationships among time series sets using state space methods.

Finally, care should be taken in any discussion of causality and especially in discussions of time series causality.  We have made many statements about failure to agree with ``intuition'' in this paper.  While some authors argue that any discussion of causality will necessarily involve appeals to intuition \cite{Pearl2000}, the possibility of intuition failing cannot be ignored completely.  

Consider the RL circuit example of Section \ref{sec:rlcirc}.  The intuitive definition of causality was motivated by an example of the experimenter physically manipulating a voltage source to create the $V$ and $I$ times series.  Suppose instead that two such experiments where conducted in isolation: one with an experimenter, Alice, physically manipulating a voltage source and measuring the current to create the $V$ and $I$ time series (call this set $\mathbf{VI}$), and another, different experiment with an experimenter, Bob, physically manipulating a current source and measuring the voltage to create the $V$ and $I$ time series (call this set $\mathbf{IV}$).  Both $\mathbf{VI}$ and $\mathbf{IV}$ are handed to a third party, Charlie, who has no {\em a priori} knowledge of how the time series are created.

Intuition for Alice is $V$ causes $I$ and she believes $\mathbf{VI}$ supports that conclusion.  Likewise, Bob believes $\mathbf{IV}$ supports his intuition that $I$ causes $V$.  Charlie, however, must rely on time series analysis alone to determine the causality in the system.  The argument we present here is not that CCM causality is insufficient because it does not provide Charlie with a definitive answer (which it does not).  Such a task is difficult and may not even be possible with time series analysis alone \cite{Pearl2000}.  The main problem is that the CCM method, as it has been explored in this work, is inconsistent.  Any method Charlie uses must be consistent if it is to be useful.  Neither Alice nor Bob would change their causality conclusions if they changed their respective input frequencies (i.e., $\Omega$ in Eqn.\ \ref{eqn:it}).  However, if Charlie used the CCM method, his causality conclusions would depend on the frequency of the signal controlled by Alice (as seen in Fig.\ \ref{fig:Av}).  Thus, CCM causality would not be a {\em consistent} tool for Charlie.  PAI was shown to give consistent results for the considered examples but does not address the ambiguity identified in this example.

\bibliography{main}

\end{document}